\documentclass[conference,a4paper]{IEEEtran}
\usepackage{verbatim}
\usepackage{cite}
\usepackage{amsmath}
\usepackage{amsfonts}
\usepackage{url}
\usepackage{graphicx}
\usepackage{subfigure}
\usepackage{caption}
 \usepackage{epstopdf}
\usepackage{multirow}
\usepackage{color,xcolor}
\usepackage{CJK}
\usepackage{comment}
\usepackage{booktabs}
\usepackage{subfigure}

\IEEEoverridecommandlockouts
\IEEEpubid{\makebox[\columnwidth]{978-1-6654-7592-1/22/\$31.00~\copyright~2024 IEEE \hfill} \hspace{\columnsep}\makebox[\columnwidth]{ }}    

\begin{document}
%
\title{Differentiable Low-computation Global Correlation Loss for Monotonicity Evaluation in Quality Assessment}

\author{\IEEEauthorblockN{Yipeng Liu}
\IEEEauthorblockA{\textit{Cooperative MediaNet Innovation Center} \\
\textit{Shanghai Jiao Tong University} \\
Shanghai, China \\
liuyipeng@sjtu.edu.cn}
\and
\IEEEauthorblockN{Qi Yang}
\IEEEauthorblockA{\textit{School of Science and Engineering} \\
\textit{University of Missouri - Kansas City} \\
Kansas, USA \\
littlleempty@gmail.com}
\and
\IEEEauthorblockN{Yiling Xu}
\IEEEauthorblockA{\textit{Cooperative MediaNet Innovation Center} \\
\textit{Shanghai Jiao Tong University} \\
Shanghai, China \\
yl.xu@sjtu.edu.cn}

}

\maketitle

\begin{abstract}

\par In this paper, we propose a global monotonicity consistency training strategy for quality assessment, which includes a differentiable, low-computation monotonicity evaluation loss function and a global perception training mechanism. Specifically, unlike conventional ranking loss and linear programming approaches that indirectly implement the Spearman rank-order correlation coefficient (SROCC) function, our method directly converts SROCC into a loss function by making the sorting operation within SROCC differentiable and functional. Furthermore, to mitigate the discrepancies between batch optimization during network training and global evaluation of SROCC, we introduce a memory bank mechanism. This mechanism stores gradient-free predicted results from previous batches and uses them in the current batch's training to prevent abrupt gradient changes. We evaluate the performance of the proposed method on both images and point clouds quality assessment tasks, demonstrating performance gains in both cases.


\end{abstract}


%
\IEEEpeerreviewmaketitle

\section{Introduction}
\label{sec:introduction}

\par Quality assessment plays a crucial role in various visual tasks, serving as an essential measure of performance and reliability across multiple domains \cite{Chen2021driving, Cheng2020medicalPC}. Accurate quality assessment ensures the effectiveness of image and video processing algorithms, enhances the realism of augmented and virtual reality applications, and supports the precision of computer vision systems employed in areas such as autonomous driving, medical imaging, and surveillance \cite{lim2020Augmented,ChenLFGW:20,Rusu2011PCL}. By providing objective metrics, quality assessment enables the optimization of visual content, ultimately improving user experience and operational efficiency. The significance of robust quality assessment methodologies is obvious, as they underpin advancements in visual technology and contribute to the broader field of visual perception research \cite{Mantiuk2012QAsurvey}.

Quality assessment metrics can be categorized into full-reference (FR), reduced-reference (RR) and no-reference (NR) types. Among these, NR metrics are garnering increased attention from researchers due to the challenges associated with obtaining reference data, particularly for evaluating the quality of media captured in the uncontrolled wild \cite{Yang2019DNNIQA,Liu2022LSPCQA,liu2022VCIP}. Numerical errors, such as mean square error (MSE) and Pearson linear correlation coefficient (PLCC), are commonly used to assess the performance of quality metrics. Consequently, MSE and PLCC are often employed as loss functions to train NR models. However, in addition to numerical errors, monotonicity indices, such as Spearman rank-order correlation coefficient (SROCC), are also important for evaluating the performance of quality assessment metrics. Optimizing SROCC directly is challenging because it involves a non-differentiable sorting operation.

Considering the critical role of monotonicity measurement in quality assessment, researchers persist in incorporating it into objective functions. One approach involves adopting the margin loss to emphasize ranking relations within sample pairs. For example, \cite{Chen2024GMCIQA,Zhang2022MMPCQA,Sun2022UGC,zeng2018novel,Liu2017RankIQA} derive quality ranking information through pairwise comparisons of samples in each batch. Another approach seeks alternatives to the non-differentiable sorting operation. For example, \cite{liu2024D3PCQA,Yang2024Attack} implement SROCC as a loss function by using a constrained linear program to achieve sorting. However, there is a distinction between pairwise ranking and global monotonicity measurement, and using pairwise comparison and linear programming for approximate sorting is computationally intensive.

In this paper, we propose global monotonicity optimization to address the aforementioned issues. The proposed approach includes a lightweight, differentiable SROCC loss function that derives ranking results using a cumulative Heaviside step function, approximated by curve functions to achieve differentiability. Additionally, a memory bank mechanism records and updates predicted quality results from previous batches as gradient-free scalars. The results collected from previous batches are involved in following batch optimization, which can decouple the isolation effect incurred by batch size and achieve global consistency. To demonstrate the effectiveness of the proposed method, we use image (IQA) and point cloud quality assessment (PCQA) tasks as a representation. Specifically, we reproduce two NR image and two NR point cloud quality metrics, and introduce the proposed strategy into their metric training procedure. The experiment results show that the proposed strategy can improve the metric performance without changing the metric framework.

\par The rest of this paper is organized as follows: Section \ref{sec:relatedwork} reviews the NR quality assessment of images and point clouds, and the loss functions used during training; Section \ref{sec:preliminary} introduces two typical existing ideas for monotonicity optimization; Section \ref{sec:methodology} presents the proposed global monotonicity optimization; Section \ref{sec:experiment} gives the experimental results; the conclusion is drawn in Section \ref{sec:conclusion}.

\section{Related Work}
\label{sec:relatedwork}

\subsection{No-Reference Image Quality Assessment (NR-IQA)}

No-reference image quality assessment (NR-IQA) has garnered significant attention due to the practical need for evaluating image quality without reference images. Traditional NR-IQA methods often employ handcrafted features to capture image distortions, such as natural scene statistics \cite{mittal2012no,Zhang2015ILNIQE,Xu2016HOSA}. With the advent of deep learning, learning-based NR-IQA methods have shown superior performance. Convolutional neural networks (CNNs) have been used to learn quality-relevant features directly from data \cite{Kim2017BIECON,Zhang2020DBCNN}. \cite{Li2019SFA} focused on blur-dominated distortions to exclude the impacts of distortion-type variations. \cite{Kim2019deepmap1,Lin2018CVPR,Pan2018CVPR} proposed to learn error maps of distorted
images to guide quality prediction. \cite{Liu2017RankIQA,Ma2017dipIQ} improved the utilization of training data by leveraging quality rank information. \cite{Su2020HyperIQA} separated the features into low-level and high-level features and transformed the latter to redirect the former. In the final analysis, all these approaches leverage large annotated datasets to minimize the distance between predicted and ground truth quality scores.

\subsection{No-Reference Point Cloud Quality Assessment (NR-PCQA)}
No-reference point cloud quality assessment (NR-PCQA) is a relatively emerging field compared to NR-IQA. Point clouds, representing 3D spatial data, present unique challenges for quality assessment due to their irregular structure and high dimensionality. \cite{Tao2021PMBVQA, Liu2021PQANet} conducted NR-PCQA by extracting features from 2D projections. \cite{Yang2021ITPCQA} proposed to leverage prior knowledge of natural images in the PCQA task. \cite{Fan2022Video,Zhang2022Video} applied video quality assessment metrics into integrated videos of point cloud projections. \cite{Liu2022LSPCQA} and \cite{shan2022GPANet} attempted to input raw 3D data into the neural network through sparse convolution and graph convolution. \cite{Zhang2022MMPCQA} leveraged both point cloud projection and raw 3D data to extract integrated features. Similar to NR-IQA, NR-PCQA also leverages large-scale training data to reduce the difference between predicted and ground truth quality scores.

\subsection{Loss Function Design in Model Training}

The loss function to make predicted quality scores conform to subjective perception serves as the core in quality assessment. Common loss functions used in NR-IQA and NR-PCQA include MSE and PLCC to reduce numerical differences between predicted and ground truth quality scores. However, these loss functions which primarily focus on numerical accuracy may not fully capture perceptual quality. To address this, researchers have explored loss functions that incorporate monotonicity constraints. Considering that the SROCC measures relative ordering of quality levels but is non-differentiable, \cite{Chen2024GMCIQA,Zhang2022MMPCQA,Sun2022UGC,zeng2018novel,Liu2017RankIQA} proposed using margin loss as an alternative to emphasize pairwise rank correlation, and \cite{liu2024D3PCQA,Yang2024Attack} introduced constrained linear programming to approximate the non-differentiable sorting operation in SROCC. However, there is a difference between small-scale measurements within a batch and global monotonicity measurement results, and utilizing pairwise comparisons and linear programming for approximate sorting is computationally intensive.


\section{Preliminary}
\label{sec:preliminary}

The monotonicity consistency performance of metrics can be measured by SROCC, which is defined as follows:
\begin{align}
SROCC\left( {{q},{{\hat q}}} \right){\rm{ }} = 1 - \frac{{6\sum\limits_{i = 1}^L {{{({m_i} - {n_i})}^2}} }}{{L({L^2} - 1)}},
\end{align}
where $q$ is the true MOS, ${\hat q}$ is the predicted quality score, $L$ is the number of distorted point clouds, $m_i$ is the rank of $q_i$ in the MOS, and $n_i$ is the rank of ${\hat q}_i$ in the predicted quality scores. SROCC can also be computed from Pearson linear correlation coefficient (PLCC):
\begin{align}
SROCC\left( {{q},{{\hat q}}} \right){\rm{ }} = {\rm{ }}PLCC\left( {R\left( {{q}} \right),R\left( {{{\hat q}}} \right)} \right),
\end{align}
where $R$ represents the sorting operation. And PLCC is defined as
\begin{align}
\label{eq15}
PLCC\left( {{q},{{\hat q}}} \right){\rm{ }} = \frac{{\sum\limits_{i=1}^{L} {({q_i} - {q_m})({{\hat q}_i} - {{\hat q}_m})} }}{{\sqrt {\sum\limits_{i=1}^{L} {{{({q_i} - {q_m})}^2}} } \sqrt {\sum\limits_{i=1}^{L} {{{({{\hat q}_i} - {{\hat q}_m})}^2}} } }},
\end{align}
where $q_m$ and ${\hat q}_m$ are the arithmetic mean of MOS $q$ and predicted scores ${\hat q}$.


As illustrated in Section \ref{sec:introduction}, to handle the non-differentiable sorting operation in SROCC, one approach is to adopt the margin loss to obtain ranking relations within sample pairs. The typical example is as follows
\begin{align}
\label{eq6}
\mathcal{L}_{\text {rank }}=\sum\limits_{i = 1}^N \sum\limits_{j = 1}^N \max \left(0,\left|\hat{q}_i-\hat{q}_j\right|-e\left(\hat{q}_i, \hat{q}_j\right) \cdot\left(q_i-q_j\right)\right),
\end{align}
where
\begin{align}
e\left(\hat{q}_i, \hat{q}_j\right)=\left\{\begin{array}{r}1, \hat{q}_i \geq \hat{q_j} \\ -1, \hat{q}_i<\hat{q_j}\end{array}\right..
\end{align}

Another approach approximates the non-differentiable sorting operation with linear programming. The sorting results can be obtained through 
\begin{align}
R^\beta(x)=\underset{z \in {conv}[N]}{\arg \min } \frac{1}{2}\left\|z+\frac{x}{\beta}\right\|^2,
\end{align}
where $\beta$ is a hyperparameter. $conv[N]$ is a convex polytope whose vertices correspond to permutations in $[N]$. The notation $[N]$ represents all permutations
of ${1, \dots , N}$ which contain $N!$ vertices. $R^\beta(x)$ calculates the Euclidean projection of $x/\beta$ on to $conv[N]$. If the projection of $x/\beta$ lies outside $conv[N]$, $R^\beta(x)$ computes the
nearest vertex of $conv[N]$ to the projection.


\section{Methodology}
\label{sec:methodology}

In this work, we try to optimize SROCC directly during training and meanwhile maintain its global consistency destroyed by batch optimization. Specifically, we define the sorting results obtained from cumulative Heaviside step functions, and then apply the curve function to obtain the approximate results. Besides, we propose a memory bank mechanism which records and updates predicted quality results from previous batches as gradient-free scalars to help the current batch's optimization achieve global consistency.

\subsection{Differentiable Approximation}

In this work, to achieve a differentiable sorting operation, we define $R$ obtained from
\begin{align}
R(x)=\sum_{x_i \in x^{(k)}} H(x-x_i),
\end{align}
where $H(x)$ is the Heaviside step function. In this way, the sorting operation of sample $x$ is represented by the number of other samples whose quality is exceeded by the sample $x$. Then we introduce a curve function to achieve its differentiability denoted as 
\begin{align}
R^{'}(x)=\sum_{x_i \in x^{(k)}} \frac{1+\tanh({k(x-x_i)})}{2},
\end{align}
where $k$ controls the steepness of the fitting function curve.

\begin{figure}[htbp]
	\centering
	\includegraphics[width=0.8\linewidth]{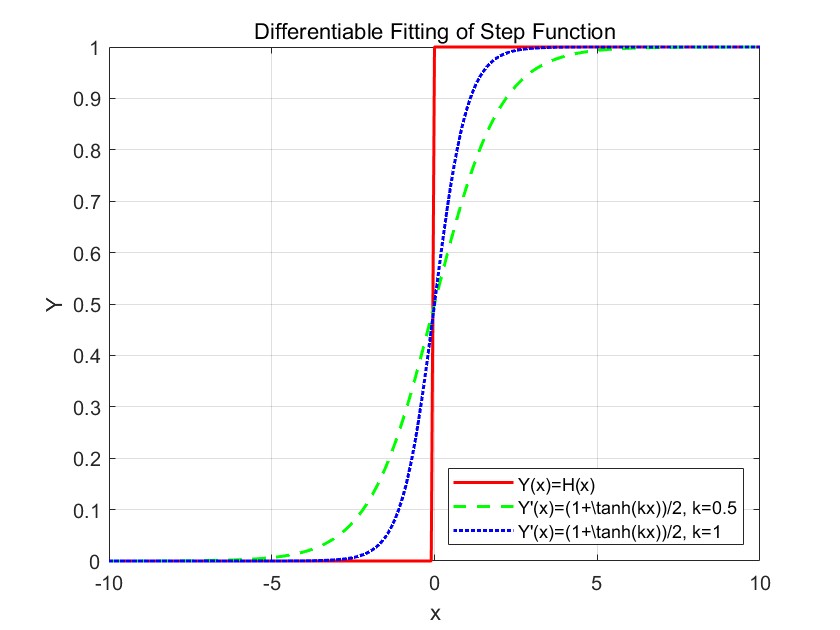}
	\caption{Differentiable fitting of step function. $k$ controls the steepness of the fitting function curve.}
	\label{reference}
\end{figure}


The gradient of the new sorting function to $x$ is formulated as 
\begin{align}
\frac{{\partial R^{'}(x)}}{{\partial x}} = \frac{k}{2} \sum_{x_i \in x^{(k)}}\left(1-\tanh ^2\left(k\left(x-x_i\right)\right)\right),
\end{align}
which exhibits continuous differentiability suitable for network training.

As a result, the monotonicity can be optimized through
\begin{align}
\label{eq10}
{\mathcal{L}_{Mono}} =  - \frac{{\sum\limits_{k = 1}^K {\left( {{r_k} - {r_m}} \right)\left( {{{\hat r}_k} - {{\hat r}_m}} \right)} }}{{\sqrt {\sum\limits_{k = 1}^K {{{\left( {{r_k} - {r_m}} \right)}^2}} } \sqrt {\sum\limits_{k = 1}^K {{{\left( {{{\hat r}_k} - {{\hat r}_m}} \right)}^2}} } }},
\end{align}
where $r_m$ and ${\hat r}_m$ are the arithmetic mean of sorting results for MOS $q$ and predicted quality scores ${\hat q}$, and the sorting results for $q_k$ and ${\hat q_k}$ are obtained through
\begin{align}
{r_k} = \sum_{q_k \in q^{(k)}} \frac{1+\tanh({k(q-q_k)})}{2} ,
\end{align}
and
\begin{align}
{\hat r_k} = \sum_{{\hat q_k} \in {\hat q^{(k)}}} \frac{1+\tanh({k({\hat q}-{\hat q_k})})}{2}.
\end{align}

Given the sample number $K$, the computational complexity of the proposed monotonicity consistency loss is $O(K)$, while the complexity of the ranking loss in Eq. \eqref{eq6} is $O(K^2)$.

\subsection{Memory Bank Mechanism}

To mitigate the destruction of global consistency by batch optimization, we establish two memory banks of predicted/ground truth quality scores in the form of dictionaries. The monotonicity consistency loss is optimized using the two dictionaries. The queries of the dictionaries are defined as the file names, while the keys are defined as the predicted or ground truth quality scores. During each batch, the corresponding terms in the two dictionaries are updated based on the output results of the network and the loading labels of the dataloader respectively. In the dictionary of predicted quality scores, only the training data of the current batch has gradients, while the training data from previous batches is scalars. And in the dictionary of ground truth MOS, the true values are continuously recorded as scalars. This operation does not introduce any additional gradients, thereby exerting minor impact on backpropagation speed. 

\begin{figure}[htbp]
\centering
\includegraphics[width=0.8\linewidth]{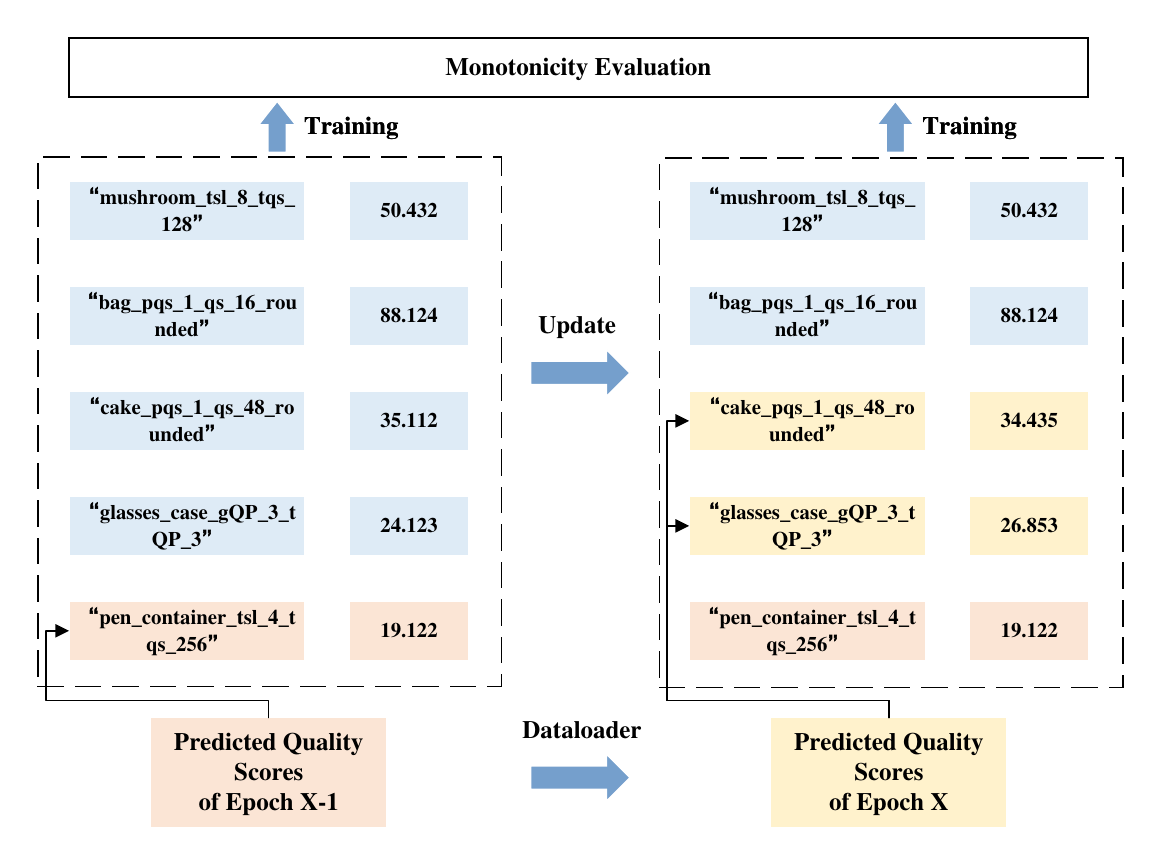}
\caption{Memory mechanism of predicted quality scores for global monotonicity evaluation. In each batch, the output quality scores of the network are recorded/updated in the memory bank. When optimizing the global monotonicity loss, the current and previous batch results are input together.}
\label{fig:memory}
\end{figure}

Eq. \eqref{eq10} is optimized using two dictionaries as MOS $q$ and predicted scores ${\hat q}$ respectively. The introduced memory banks decouple the batch size from the dictionary size which can be much larger than the regular batch size. And the length of memory banks can be controlled by retaining the results of the last $N$ epochs. As a result, during training, the model can perceive the approximate quality scores of more samples (the predicted quality scores in the previous mini-batches), and promote the global content consistency.

\section{Experiments}
\label{sec:experiment}

\par In this section, we evaluate the performance of the proposed global monotonicity optimization in quality assessment for both images and point clouds. 

\subsection{Datasets}

For images, we evaluate the performance of the proposed global monotonicity optimization on CSIQ \cite{Larson2010CSIQ} and LIVE \cite{Sheikh2006LIVE}. 

For point clouds, we evaluate the performance of the proposed global monotonicity optimization on SJTU-PCQA~\cite{Yang2020TMM3DTO2D} and WPC~\cite{Su2019WPC,Liu2022WPC}. 

\subsection{Implementation Details}

We evaluate the performance of the proposed global monotonicity optimization by adding the proposed loss function into the objective functions and introducing the memory bank mechanism into the training procedure of the learning-based quality assessment methods. The experiments are performed using PyTorch on NVIDIA 3090 GPUs. The size of images and point cloud projections, batch size, and default training duration are the same as in the reproduced metrics.

To alleviate the content biases, 10-fold cross-validation is adopted. For CSIQ and LIVE, the datasets are split into the training-testing sets with a ratio of 4:1 according to the reference images for each fold, and the median performance in the testing sets is recorded \cite{Su2020HyperIQA}. For SJTU-PCQA and WPC, the datasets are split with a ratio of 7:2 and 4:1 respectively based on the reference point clouds, and the average performance is recorded \cite{Liu2021PQANet}.


\subsection{Performance Gain}

To evaluate the performance of the proposed global monotonicity optimization, we reproduce DBCNN \cite{Zhang2020DBCNN} and HyperIQA \cite{Su2020HyperIQA} for images, and ResSCNN \cite{Liu2022LSPCQA} and PQA-Net \cite{Liu2021PQANet} for point clouds, and applying the proposed global monotonicity optimization into their training procedure. Among the PCQA metrics, PQA-Net is conducted based on point cloud projections, and ResSCNN is implemented based on raw 3D point cloud data (additional downsampling into unified 2,500 points is carried out to accelerate training speed, denoted as ResSCNN(D)). Besides, we also list the performance of other learning-based quality assessment methods, such as BRISQUE \cite{mittal2012no}, ILNIQE \cite{Zhang2015ILNIQE}, HOSA \cite{Xu2016HOSA}, BIECON \cite{Kim2017BIECON}, WaDIQaM \cite{Bosse2018WaDIQaM}, SFA \cite{Li2019SFA} and PQR \cite{Zeng2017PQR} for images, and GPANet \cite{shan2022GPANet}, MM-PCQA \cite{Zhang2022MMPCQA} and IT-PCQA \cite{Yang2021ITPCQA} for point clouds, to exhibit the overall performance. The performance gain for quality assessment is shown in Table \ref{tab:overallIQA} for images and in Table \ref{tab:overallPCQA} for point clouds.

\begin{table}[htbp]
  \centering
  \caption{Overall performance gain for image quality assessment.}
    \begin{tabular}{l|cc|cc}
    \hline
    \multirow{2}{*}{Method} & \multicolumn{2}{c|}{CSIQ} & \multicolumn{2}{c}{LIVE} \\
\cline{2-5}          & PLCC  & SROCC & PLCC  & SROCC \\
    \hline
    BRISQUE \cite{mittal2012no} & 0.829  & 0.746  & 0.935  & 0.939  \\
    ILNIQE \cite{Zhang2015ILNIQE} & 0.808  & 0.806  & 0.865  & 0.902  \\
    HOSA \cite{Xu2016HOSA} & 0.823  & 0.741  & 0.947  & 0.946  \\
    BIECON \cite{Kim2017BIECON} & 0.823  & 0.815  & 0.962  & 0.961  \\
    WaDIQaM \cite{Bosse2018WaDIQaM} & 0.973  & 0.955  & 0.963  & 0.954  \\
    SFA \cite{Li2019SFA}  & 0.818  & 0.796  & 0.895  & 0.883  \\
    PQR \cite{Zeng2017PQR}  & 0.901  & 0.873  & 0.971  & 0.965  \\
    \hline
    DBCNN \cite{Zhang2020DBCNN} & 0.949  & 0.971  & 0.965  & 0.967  \\
    DBCNN (/w ours) & \textbf{0.953}  & \textbf{0.972}  & \textbf{0.967}  & \textbf{0.967}  \\
    \hline
    HyperIQA \cite{Su2020HyperIQA} & \textbf{0.937}  & \textbf{0.922}  & 0.970  & 0.969  \\
    HyperIQA (/w ours) & 0.935  & 0.914  & \textbf{0.972}  & \textbf{0.971}  \\
    \hline
    \end{tabular}%
  \label{tab:overallIQA}%
\end{table}%

\begin{table}[htbp]
  \centering
  \caption{Overall performance gain for point cloud quality assessment.}
    \begin{tabular}{l|cc|cc}
    \hline
    \multirow{2}{*}{Method} & \multicolumn{2}{c|}{SJTU-PCQA} & \multicolumn{2}{c}{WPC} \\
\cline{2-5}          & PLCC  & SROCC & PLCC  & SROCC \\
    \hline
    GPA-Net \cite{shan2022GPANet} & 0.864  & 0.855  & 0.722  & 0.704  \\
    MM-PCQA \cite{Zhang2022MMPCQA} & 0.924  & 0.902  & 0.840  & 0.825  \\
    IT-PCQA \cite{Yang2021ITPCQA} & 0.609  & 0.597  & 0.570  & 0.552  \\
    \hline
    ResSCNN(D) \cite{Liu2022LSPCQA} & 0.665  & 0.668  & 0.446  & 0.434  \\
    ResSCNN(D) (/w ours) & \textbf{0.672}  & \textbf{0.679}  & \textbf{0.453}  & \textbf{0.447}  \\
    \hline
    PQA-Net \cite{Liu2021PQANet} & 0.859  & 0.836  & 0.718  & 0.703  \\
    PQA-Net (/w ours) & \textbf{0.860}  & \textbf{0.847}  & \textbf{0.723}  & \textbf{0.715}  \\
    \hline
    \end{tabular}%
  \label{tab:overallPCQA}%
\end{table}%

We can see that the proposed global monotonicity optimization exhibits performance gain under most testing conditions, demonstrating its effectiveness.

\subsection{Ablation Study}

To evaluate the performance of the proposed differentiable SROCC loss and memory bank mechanism respectively, we conduct the evaluation experiments on the LIVE and WPC datasets based on HyperIQA \cite{Su2020HyperIQA} and PQA-Net \cite{Liu2021PQANet}. Specifically, we compare the initial performance, the performance with only differentiable SROCC loss (denoted as (/w $\mathcal{L}_{Mono}$)), and the performance with both differentiable SROCC loss and memory bank mechanism (i.e. the final performance, denoted as (/w $\mathcal{L}_{Mono}$ + M)). The results are shown in Table \ref{tab:ablationIQA} and Table \ref{tab:ablationPCQA}.

\begin{table}[htbp]
  \centering
  \caption{Performance gain of different modules on LIVE.}
    \begin{tabular}{l|cc}
    \hline
          & PLCC  & \multicolumn{1}{c}{SROCC} \\
    \hline
    HyperIQA & 0.970  & 0.969  \\
    HyperIQA (/w $\mathcal{L}_{Mono}$) & 0.972 & 0.970 \\
    HyperIQA (/w $\mathcal{L}_{Mono}$ + M) & 0.972  & 0.971  \\
    \hline
    \end{tabular}%
  \label{tab:ablationIQA}%
\end{table}%

\begin{table}[htbp]
  \centering
  \caption{Performance gain of different modules on WPC.}
    \begin{tabular}{l|cc}
    \hline
          & PLCC  & \multicolumn{1}{c}{SROCC} \\
    \hline
    PQA-Net & 0.718  & 0.703  \\
    PQA-Net (/w $\mathcal{L}_{Mono}$) & 0.721 & 0.711 \\
    PQA-Net (/w $\mathcal{L}_{Mono}$ + M) & 0.723  & 0.715  \\
    \hline
    \end{tabular}%
  \label{tab:ablationPCQA}%
\end{table}%

We can see that both differentiable SROCC loss and memory bank mechanism exhibit performance gain for quality assessment.

\section{Conclusion}
\label{sec:conclusion}

\par In this paper, we propose a global monotonicity consistency training strategy for quality assessment, which includes a differentiable, low-computation monotonicity evaluation loss function and a global perception training mechanism. We evaluate the performance of the proposed method for quality assessment on both images and point clouds, demonstrating performance gains in both cases.

\section{Acknowledgement}

This paper is supported in part by National Natural Science Foundation of China (62371290, U20A20185), the Fundamental Research Funds for the Central Universities of China, and 111 project (BP0719010). The corresponding author is Yiling Xu(e-mail: yl.xu@sjtu.edu.cn).

\bibliography{manuscript}

\begin{thebibliography}{10}
\providecommand{\url}[1]{#1}
\csname url@samestyle\endcsname
\providecommand{\newblock}{\relax}
\providecommand{\bibinfo}[2]{#2}
\providecommand{\BIBentrySTDinterwordspacing}{\spaceskip=0pt\relax}
\providecommand{\BIBentryALTinterwordstretchfactor}{4}
\providecommand{\BIBentryALTinterwordspacing}{\spaceskip=\fontdimen2\font plus
\BIBentryALTinterwordstretchfactor\fontdimen3\font minus \fontdimen4\font\relax}
\providecommand{\BIBforeignlanguage}[2]{{%
\expandafter\ifx\csname l@#1\endcsname\relax
\typeout{** WARNING: IEEEtran.bst: No hyphenation pattern has been}%
\typeout{** loaded for the language `#1'. Using the pattern for}%
\typeout{** the default language instead.}%
\else
\language=\csname l@#1\endcsname
\fi
#2}}
\providecommand{\BIBdecl}{\relax}
\BIBdecl

\bibitem{Chen2021driving}
S.~Chen, B.~Liu, C.~Feng \emph{et~al.}, ``{3D} point cloud processing and learning for autonomous driving: Impacting map creation, localization, and perception,'' \emph{IEEE Signal Processing Magazine}, vol.~38, no.~1, pp. 68--86, 2021.

\bibitem{Cheng2020medicalPC}
Q.~Cheng, P.~Sun, C.~Yang \emph{et~al.}, ``A morphing-based {3D} point cloud reconstruction framework for medical image processing,'' \emph{Computer Methods and Programs in Biomedicine}, vol. 193, p. 105495, 2020.

\bibitem{lim2020Augmented}
S.~Lim, M.~Shin, and J.~Paik, ``Point cloud generation using deep local features for augmented and mixed reality contents,'' in \emph{IEEE International Conference on Consumer Electronics}, 2020, pp. 1--3.

\bibitem{ChenLFGW:20}
S.~Chen, B.~Liu, C.~Feng \emph{et~al.}, ``{3D} point cloud processing and learning for autonomous driving: Impacting map creation, localization, and perception,'' \emph{IEEE Signal Processing Magazine}, vol.~38, no.~1, pp. 68--86, 2021.

\bibitem{Rusu2011PCL}
R.~B. Rusu and S.~Cousins, ``3d is here: Point cloud library ({PCL}),'' in \emph{IEEE International Conference on Robotics and Automation}, 2011, pp. 1--4.

\bibitem{Mantiuk2012QAsurvey}
R.~K. Mantiuk, A.~Tomaszewska, and R.~Mantiuk, ``Comparison of four subjective methods for image quality assessment,'' \emph{Computer Graphics Forum}, vol.~31, no.~8, pp. 2478--2491, 2012.

\bibitem{Yang2019DNNIQA}
X.~Yang, F.~Li, and H.~Liu, ``A survey of {DNN} methods for blind image quality assessment,'' \emph{IEEE Access}, vol.~7, pp. 123\,788--123\,806, 2019.

\bibitem{Liu2022LSPCQA}
Y.~Liu, Q.~Yang, Y.~Xu, and L.~Yang, ``Point cloud quality assessment: Dataset construction and learning-based no-reference metric,'' \emph{ACM Transactions on Multimedia Computing Communications and Applications}, vol.~19, no.~2s, pp. 1--26, 2023.

\bibitem{liu2022VCIP}
Y.~Liu, Q.~Yang, and Y.~Xu, ``Reduced reference quality assessment for point cloud compression,'' in \emph{IEEE International Conference on Visual Communications and Image Processing}, 2022, pp. 1--5.

\bibitem{Chen2024GMCIQA}
Z.~Chen, J.~Wang, B.~Li \emph{et~al.}, ``{GMC-IQA}: Exploiting global-correlation and mean-opinion consistency for no-reference image quality assessment,'' \emph{arXiv preprint arXiv:2401.10511}, 2024.

\bibitem{Zhang2022MMPCQA}
Z.~Zhang, W.~Sun, X.~Min \emph{et~al.}, ``{MM-PCQA}: Multi-modal learning for no-reference point cloud quality assessment,'' in \emph{International Joint Conference on Artificial Intelligence}, 2023, pp. 1--1.

\bibitem{Sun2022UGC}
W.~Sun, X.~Min, W.~Lu, and G.~Zhai, ``A deep learning based no-reference quality assessment model for ugc videos,'' in \emph{ACM International Conference on Multimedia}, 2022, p. 856–865.

\bibitem{zeng2018novel}
X.~Zeng and H.~Lu, ``Novel no-reference image quality assessment metric based on joint relative features,'' \emph{IEEE Transactions on Image Processing}, vol.~27, no.~8, pp. 4010--4022, 2018.

\bibitem{Liu2017RankIQA}
X.~Liu, J.~van~de Weijer, and A.~D. Bagdanov, ``{RankIQA}: Learning from rankings for no-reference image quality assessment,'' in \emph{IEEE International Conference on Computer Vision}, 2017.

\bibitem{liu2024D3PCQA}
Y.~Liu, Q.~Yang, Y.~Zhang \emph{et~al.}, ``{Once-Training-All-Fine}: No-reference point cloud quality assessment via domain-relevance degradation description,'' \emph{arXiv preprint arXiv:2307.01567}, 2023.

\bibitem{Yang2024Attack}
C.~Yang, Y.~Liu, D.~Li \emph{et~al.}, ``Beyond score changes: Adversarial attack on no-reference image quality assessment from two perspectives,'' \emph{arXiv preprint arXiv:2404.13277}, 2024.

\bibitem{mittal2012no}
A.~Mittal, A.~K. Moorthy, and A.~C. Bovik, ``No-reference image quality assessment in the spatial domain,'' \emph{IEEE Transactions on Image Processing}, vol.~21, no.~12, pp. 4695--4708, 2012.

\bibitem{Zhang2015ILNIQE}
L.~Zhang, L.~Zhang, and A.~C. Bovik, ``A feature-enriched completely blind image quality evaluator,'' \emph{IEEE Transactions on Image Processing}, vol.~24, no.~8, pp. 2579--2591, 2015.

\bibitem{Xu2016HOSA}
J.~Xu, P.~Ye, Q.~Li \emph{et~al.}, ``Blind image quality assessment based on high order statistics aggregation,'' \emph{IEEE Transactions on Image Processing}, vol.~25, no.~9, pp. 4444--4457, 2016.

\bibitem{Kim2017BIECON}
J.~Kim and S.~Lee, ``Fully deep blind image quality predictor,'' \emph{IEEE Journal of Selected Topics in Signal Processing}, vol.~11, no.~1, pp. 206--220, 2017.

\bibitem{Zhang2020DBCNN}
W.~Zhang, K.~Ma, J.~Yan \emph{et~al.}, ``Blind image quality assessment using a deep bilinear convolutional neural network,'' \emph{IEEE Transactions on Circuits and Systems for Video Technology}, vol.~30, no.~1, pp. 36--47, 2020.

\bibitem{Li2019SFA}
D.~Li, T.~Jiang, W.~Lin, and M.~Jiang, ``Which has better visual quality: The clear blue sky or a blurry animal?'' \emph{IEEE Transactions on Multimedia}, vol.~21, no.~5, pp. 1221--1234, 2019.

\bibitem{Kim2019deepmap1}
J.~Kim, A.-D. Nguyen, and S.~Lee, ``Deep cnn-based blind image quality predictor,'' \emph{IEEE Transactions on Neural Networks and Learning Systems}, vol.~30, no.~1, pp. 11--24, 2019.

\bibitem{Lin2018CVPR}
K.-Y. Lin and G.~Wang, ``Hallucinated-iqa: No-reference image quality assessment via adversarial learning,'' in \emph{IEEE/CVF Conference on Computer Vision and Pattern Recognition}, 2018.

\bibitem{Pan2018CVPR}
D.~Pan, P.~Shi, M.~Hou \emph{et~al.}, ``Blind predicting similar quality map for image quality assessment,'' in \emph{IEEE/CVF Conference on Computer Vision and Pattern Recognition}, 2018.

\bibitem{Ma2017dipIQ}
K.~Ma, W.~Liu, T.~Liu \emph{et~al.}, ``{dipIQ}: Blind image quality assessment by learning-to-rank discriminable image pairs,'' \emph{IEEE Transactions on Image Processing}, vol.~26, no.~8, pp. 3951--3964, 2017.

\bibitem{Su2020HyperIQA}
S.~Su, Q.~Yan, Y.~Zhu \emph{et~al.}, ``Blindly assess image quality in the wild guided by a self-adaptive hyper network,'' in \emph{IEEE/CVF Conference on Computer Vision and Pattern Recognition}, 2020.

\bibitem{Tao2021PMBVQA}
W.~Tao, G.~Jiang, Z.~Jiang, and M.~Yu, ``Point cloud projection and multi-scale feature fusion network based blind quality assessment for colored point clouds,'' in \emph{ACM International Conference on Multimedia}, 2021, p. 5266–5272.

\bibitem{Liu2021PQANet}
Q.~Liu, H.~Yuan, H.~Su \emph{et~al.}, ``{PQA-Net}: Deep no reference point cloud quality assessment via multi-view projection,'' \emph{IEEE Transactions on Circuits and Systems for Video Technology}, vol.~31, no.~12, pp. 4645--4660, 2021.

\bibitem{Yang2021ITPCQA}
Q.~Yang, Y.~Liu, S.~Chen \emph{et~al.}, ``No-reference point cloud quality assessment via domain adaptation,'' in \emph{IEEE/CVF Conference on Computer Vision and Pattern Recognition}, 2022, pp. 21\,179--21\,188.

\bibitem{Fan2022Video}
Y.~Fan, Z.~Zhang, W.~Sun \emph{et~al.}, ``A no-reference quality assessment metric for point cloud based on captured video sequences,'' in \emph{IEEE International Workshop on Multimedia Signal Processing}, 2022, pp. 1--5.

\bibitem{Zhang2022Video}
Z.~Zhang, W.~Sun, Y.~Zhu \emph{et~al.}, ``Evaluating point cloud from moving camera videos: A no-reference metric,'' \emph{arXiv preprint arXiv:2208.14085}, 2022.

\bibitem{shan2022GPANet}
Z.~Shan, Q.~Yang, R.~Ye \emph{et~al.}, ``{GPA-Net}: No-reference point cloud quality assessment with multi-task graph convolutional network,'' \emph{IEEE Transactions on Visualization and Computer Graphics}, pp. 1--13, 2023.

\bibitem{Larson2010CSIQ}
E.~C. Larson and D.~M. Chandler, ``{Most apparent distortion: full-reference image quality assessment and the role of strategy},'' \emph{Journal of Electronic Imaging}, vol.~19, no.~1, p. 011006, 2010.

\bibitem{Sheikh2006LIVE}
H.~Sheikh, M.~Sabir, and A.~Bovik, ``A statistical evaluation of recent full reference image quality assessment algorithms,'' \emph{IEEE Transactions on Image Processing}, vol.~15, no.~11, pp. 3440--3451, 2006.

\bibitem{Yang2020TMM3DTO2D}
Q.~Yang, H.~Chen, Z.~Ma \emph{et~al.}, ``Predicting the perceptual quality of point cloud: A 3d-to-2d projection-based exploration,'' \emph{IEEE Transactions on Multimedia}, vol.~23, pp. 3877--3891, 2021.

\bibitem{Su2019WPC}
H.~Su, Z.~Duanmu, W.~Liu \emph{et~al.}, ``Perceptual quality assessment of {3D} point clouds,'' in \emph{IEEE International Conference on Image Processing}, 2019, pp. 3182--3186.

\bibitem{Liu2022WPC}
Q.~Liu, H.~Su, Z.~Duanmu \emph{et~al.}, ``Perceptual quality assessment of colored {3D} point clouds,'' \emph{IEEE Transactions on Visualization and Computer Graphics}, vol.~29, no.~8, pp. 3642--3655, 2023.

\bibitem{Bosse2018WaDIQaM}
S.~Bosse, D.~Maniry, K.-R. Müller \emph{et~al.}, ``Deep neural networks for no-reference and full-reference image quality assessment,'' \emph{IEEE Transactions on Image Processing}, vol.~27, no.~1, pp. 206--219, 2018.

\bibitem{Zeng2017PQR}
H.~Zeng, L.~Zhang, and A.~C. Bovik, ``A probabilistic quality representation approach to deep blind image quality prediction,'' \emph{arXiv preprint arXiv:1708.08190}, 2017.

\end{thebibliography}
\bibliographystyle{IEEEtran}

\end{document}